\documentclass[twocolumn,showpacs]{revtex4}
\usepackage{graphicx}
\usepackage{dcolumn}
\usepackage{bm}
\usepackage{amsmath}
\usepackage{amsfonts}
\usepackage{amssymb}

\providecommand{\U}[1]{\protect\rule{.1in}{.1in}}

\begin{document}

\title{Quantum walks: decoherence and coin-flipping games}
\author{Alejandro Romanelli}
\altaffiliation{alejo@fing.edu.uy}
\author{Guzm\'an Hern\'andez}
\altaffiliation{guzmanhc@fing.edu.uy}
\affiliation{Instituto de F\'{\i}sica, Facultad de Ingenier\'{\i}a\\
Universidad de la Rep\'ublica\\
C.C. 30, C.P. 11000, Montevideo, Uruguay}
\date{\today }

\begin{abstract}
We investigate the global chirality distribution of the quantum walk
on the line when decoherence is introduced either through
simultaneous measurements of the chirality and particle position, or
as a result of broken links. The first mechanism drives the system
towards a classical diffusive behavior. This is used to build new
quantum games, similar to the spin-flip game. The second mechanism
involves two different possibilities: (a) All the quantum walk links
have the same probability of being broken. (b) Only the quantum walk
links on a half-line are affected by random breakage. In case (a)
the decoherence drives the system to a classical Markov process,
whose master equation is equivalent to the dynamical equation of the
quantum density matrix. This is not the case in (b) where the
asymptotic global chirality distribution unexpectedly maintains some
dependence with the initial condition. Explicit analytical equations
are obtained for all cases.
\end{abstract}

\maketitle

\section{Introduction}

In the field of quantum computation, discoveries like the Shor and
Grover algorithms \cite{Shor,Grover} have shown a superior
efficiency with regard to their classical equivalents. In this frame
the quantum walk (QW) on the line \cite{QW}, a natural
generalization of the classical random walk, is receiving permanent
attention\cite{childs,Romanelli09,Linden}. It has been used as the
basis for optimal quantum search algorithms \cite{Shenvi,Childs et}
on several topologies and more recently it has been shown that any
quantum algorithm can be reformulated in terms of a QW algorithm
\cite{childs,Lovett}, this means that the QW is universal in quantum
computation. In particular the QW has the property to spread over
the line quadratically faster than its classical analog. It remains
a challenge to use this quantum property, as well as quantum
parallelism and quantum entanglement, to increase the speed and
efficiency of the algorithms. In this line of thought, it is
important to probe further into the properties of the QW dynamics.

On the other hand the spectacular development of techniques to
manipulate atoms and photons has led to some experimental
implementations of the QW \cite{Exp} as well as several
implementation proposals \cite{ProExp}. Here the obstacle of
decoherence is present owing to imperfections and environmental
noise.

Several authors have studied the QW subjected to different types of
coin operators and sources of decoherence to analyze and verify the
principles of quantum theory as well as the passage from the quantum
to the classical world \cite{QW2,alejo4,alejo1,Brun,Brun1}. These
works have shown that the decoherence transforms the quantum
behavior into a classical diffusive behavior. In general, to obtain
this conclusion the authors focus their studies in the spreading of
the wave-function. However decoherence may also drive the system to
a more general behavior than the diffusive. \cite{Romanelli09k}.

In recent work \cite{Romanelli10} the global chirality distribution
(GCD) was defined as the distribution of the chirality independently
of position and the asymptotic behavior of the QW was investigated
focusing on the GCD without decoherence. It was shown that the GCD
has a stationary long-time limit, a result usually linked to a
Markovian process.

In the present paper, we study the asymptotic behavior of the GCD
when the system is subjected to two different sources of
decoherence. In the first place, decoherence is introduced through
simultaneous measurements of the chirality and particle position.
This treatment allows us to connect the dynamics of the QW with
certain aspects of quantum game theory \cite{Meyer}. In the second
place, the decoherence is introduced through the influence of
randomly broken links into the QW dynamics. This last mechanism may
be relevant in experimental realizations of quantum computation
based on Ising spin-$1/2$ chains in solid-state substrates
\cite{Berman}.

In the next section we present the standard QW model and the GCD
dynamics. In the third section we build the master equation for the
GCD with periodic measurement. In the fourth section we present the
coin-flipping games with the GCD. In the fifth section the GCD
dynamics with broken links is studied, and in the last section we
draw the conclusions.

\section{\textbf{The coined QW and the GCD}}

In this QW the walker moves (at discrete time steps
$t\in\mathbb{N}$) along a one-dimensional lattice of sites
$k\in\mathbb{Z}$, with a direction that depends on the state of the
coin. The global Hilbert space of the system is the tensor product
$H_{s}\otimes H_{c}$ where $H_{s}$ is the Hilbert space associated
to the motion on the line and $H_{c}$ is the chirality Hilbert
space. The one-qubit \textquotedblleft coin\textquotedblright\
subspace, $H_{c}$, is spanned by two orthogonal states
\{$|L\rangle$, $|R\rangle$\}. The spatial subspace, $H_{s}$, is
spanned by the orthogonal set of position eigenstates, $|k\rangle$.
The evolution is generated by repeated application of a composite
unitary operator $U$ which implements a coin operation followed by a
conditional shift in the position of the walker. This shift
operation entangles the coin and position of the walker. In our case
$U$ depends on a parameter $\theta$, with $\theta\in\left[
0,\pi/2\right] ,$ that defines the bias of the coin toss
($\theta=\frac{\pi}{4}$ for an unbiased, or Hadamard, coin). Then
the unitary operator $U(\theta)$ evolves the state in one time step
as
\begin{equation}
|\Psi(t+1)\rangle=U(\theta)|\Psi(t)\rangle.   \label{evolution}
\end{equation}
The state of the total system at time $t$ can be expressed as the spinor
\begin{equation}
|\Psi(t)\rangle=\sum\limits_{k=-\infty}^{\infty}\left[
\begin{array}{c}
a_{k}(t) \\
b_{k}(t)%
\end{array}
\right] |k\rangle,   \label{spinor}
\end{equation}
where the upper (lower) component is associated to the left (right)
chirality. Therefore, the QW is ruled by a unitary map, its standard
form being \cite{Romanelli09}
\begin{align}
a_{k}(t+1) & =a_{k+1}(t)\,\cos\theta\,+b_{k+1}(t)\,\sin\theta,\,  \notag \\
b_{k}(t+1) & =a_{k-1}(t)\,\sin\theta\,-b_{k-1}(t)\,\cos\theta.   \label{mapa}
\end{align}

In references \cite{Alejo1,Alejo2} it is shown how a unitary quantum
mechanical evolution can be separated into Markovian and
interference terms. This idea has been implemented recently in Ref.
\cite{Romanelli10} for the global left and right chirality
probabilities which are defined by
\begin{align}
P_{L}(t) & = \sum_{k=-\infty}^{\infty}\left\vert a_{k}(t)\right\vert ^{2},\,
\label{pl} \\
P_{R}(t) & =\sum_{k=-\infty}^{\infty}\left\vert b_{k}(t)\right\vert ^{2},
\label{pr}
\end{align}
with $P_{R}(t)+P_{L}(t)=1$.  The pair formed by $%
\left[
\begin{array}{c}
P_{L}(t) \\
P_{R}(t)%
\end{array}
\right] $ is called the global chirality distribution (GCD).
Starting from the original map Eq. (\ref{mapa})
a quantum dynamical equation for the these distributions was obtained \cite%
{Romanelli10}
\begin{align}
{\left[
\begin{array}{c}
P_{L}(t+1) \\
P_{R}(t+1)%
\end{array}
\right] } & ={\left(
\begin{array}{cc}
\cos^{2}\theta & \sin^{2}\theta \\
\sin^{2}\theta & \cos^{2}\theta%
\end{array}
\right) }\left[
\begin{array}{c}
P_{L}(t) \\
P_{R}(t)%
\end{array}
\right]  \notag \\
& +\mathrm{Re}\left[ Q(t)\right] \sin{2}\theta\left[
\begin{array}{c}
1 \\
-1%
\end{array}
\right] ,   \label{master}
\end{align}
where
\begin{equation}
Q(t)=\sum_{k=-\infty}^{\infty}a_{k}(t)b_{k}^{\ast}(t),   \label{inter}
\end{equation}
with $\mathrm{Re}(z)$ indicating the real part of $z$. The two
dimensional matrix in Eq. (\ref{master}) can be interpreted as a
transition probability matrix for a classical two dimensional random
walk. On the other hand, it is clear that $Q(t)$ accounts for the
interference. When $Q(t)$ vanishes the behavior of the GCD can be
described as a classical Markovian process.

It has been proven \cite{Romanelli10} that $Q(t)$, $P_{L}(t)$ and
$P_{R}(t)$ have a long-time limit, and their values are determined
by the initial conditions. Eq. (\ref{master}) was solved in this
limit defining
\begin{align}
\Pi_{L} & \equiv P_{L}(t\rightarrow\infty),\,  \notag \\
\Pi_{R} & \equiv P_{R}(t\rightarrow\infty),\,  \notag \\
Q_{0} & \equiv Q(t\rightarrow\infty),  \label{asym}
\end{align}
and substituting these values in Eq. (\ref{master}). The stationary
solution obtained for the GCD was
\begin{equation}
{\left[
\begin{array}{c}
\Pi_{L} \\
\Pi_{R}%
\end{array}
\right] }=\frac{1}{2}\left[
\begin{array}{c}
1+2\mathrm{Re}(Q_{0})/\tan\theta \\
1-2\mathrm{Re}(Q_{0})/\tan\theta%
\end{array}
\right] .   \label{estacio}
\end{equation}
Therefore, the dynamical evolution of the QW is unitary but the evolution of
its GCD has an asymptotic value characteristic of a diffusive behavior.

In this paper the initial state of the walker is assumed to be
sharply localized at the line origin with arbitrary chirality
\begin{equation}
|\Psi(0)\rangle=\left[
\begin{array}{c}
\cos\alpha \\
\exp i\beta\text{ }\sin\alpha%
\end{array}
\right] |0\rangle,   \label{psi0}
\end{equation}
where $\alpha\in\left[ 0,\pi\right] $ and $\beta\in\left[
0,2\pi\right] $ define a point on the unit three-dimensional Bloch
sphere. Fixing the bias of the coin toss $\theta=\pi/4$, the
analytical value of $Q_{0}$ was obtained in Ref. \cite{abals}
following the method developed by Nayak and
Vishwanath \cite{nayak}%
\begin{equation}
Q_{0}=\frac{1}{2}(1-\frac{1}{\sqrt{2}})\left[ \cos2\alpha\text{ }+\sin
2\alpha\text{ }(\cos\beta+i\sqrt{2}\sin\beta)\right] .   \label{q0entengl}
\end{equation}
Using this expression in Eq. (\ref{estacio}) the asymptotic
long-time limit of the GCD is
\begin{equation}
{\left[
\begin{array}{c}
\Pi_{L} \\
\Pi_{R}%
\end{array}
\right] }=\frac{1}{2}\left[
\begin{array}{c}
1+(1-1/\sqrt{2})\left[ \cos2\alpha\text{ +}\cos\beta\text{ }\sin 2\alpha%
\right] \\
1-(1-1/\sqrt{2})\left[ \cos2\alpha\text{ +}\cos\beta\text{ }\sin 2\alpha%
\right]%
\end{array}
\right] .   \label{parti}
\end{equation}
If we assume that the parameters $\alpha$ and $\beta$ verify
$\tan2\alpha=-1/\cos\beta$, the asymptotic solution is
$\Pi_{L}=\Pi_{R}=1/2$ and $\mathrm{Re(}Q_{0})=0$. In this case Eq.
(\ref{master}) approaches a Markov chain \cite{Cox} with two states
and the dynamics of the GDC turns into an example of dependent
Bernoulli trials in which the probabilities of success or failure at
each trial depend on the outcome of the previous trial.

\begin{figure}[h]
\begin{center}
\includegraphics[scale=0.4]{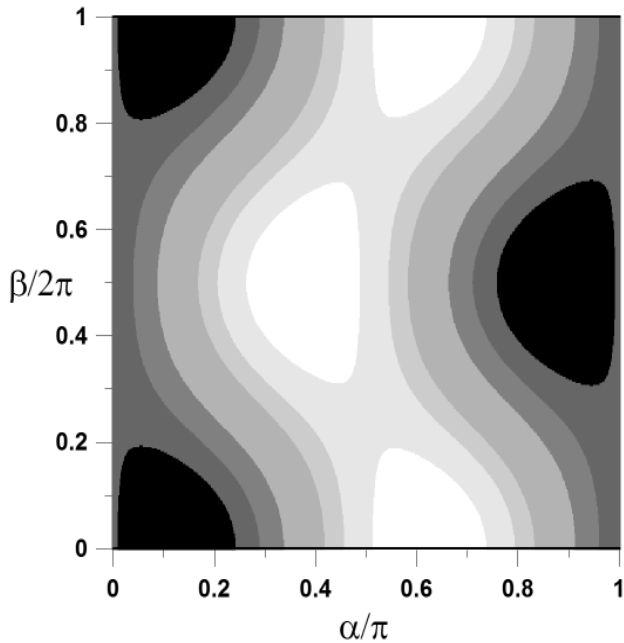}
\end{center}
\caption{The asymptotic probability $\Pi_{L}$ as a function of
dimensionless angles using Eq. (\protect\ref{parti}). Seven zones of
probability are given with different tonalities of gray. From the
black to the white zone the
intervals of probability are: (1) $\left[0.29,0.34\right]$, (2) $\left[%
0.34,0.39\right]$, (3) $\left[0.39,0.45\right] $, (4) $\left[0.45,0.52\right]$%
, (5) $\left[0.52,0.60\right]$, (6) $\left[0.60,0.65\right]$, (7) $\left[%
0.65,0.71\right]$. In section IV this figure also represents Alice's
strategic space for the game with one measurement} \label{f1}
\end{figure}

\begin{figure}[h]
\begin{center}
\includegraphics[scale=0.4]{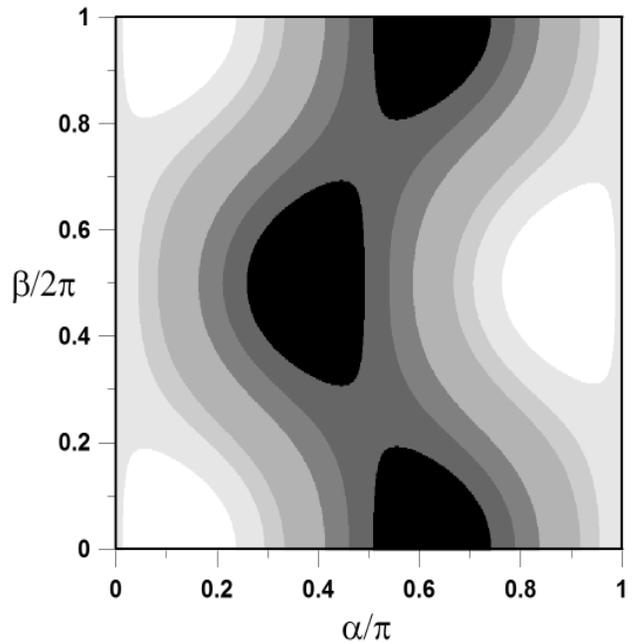}
\end{center}
\caption{The asymptotic probability $\Pi_{R}$ as a function of
dimensionless angles using Eq. (\protect\ref{parti}). Seven zones of
probability are given with different tonalities of gray. From the
black to the white zone the
intervals of probability are: (1) $\left[0.29,0.34\right]$, (2) $\left[%
0.34,0.39\right]$, (3) $\left[0.39,0.45\right] $, (4) $\left[0.45,0.52\right]$%
, (5) $\left[0.52,0.60\right]$, (6) $\left[0.60,0.65\right]$, (7) $\left[%
0.65,0.71\right]$. In section IV this figure also represents Bob's
strategic space for the game with one measurement.} \label{f2}
\end{figure}

In Figs. ~\ref{f1} and \ref{f2} we present the probabilities
$\Pi_{L}$ and $\Pi_{R}$, respectively, as functions of the
parameters $\alpha$ and $\beta$. In these figures we use $2000$ time
steps as an asymptotic temporal value. Zones with different
tonalities of gray can be appreciated representing different
occurrence probabilities. The values of $\Pi_{L}$ and $\Pi_{R}$ are
approximately restricted to the interval $\left[0.3,0.7\right]$ as a
consequence of the dependence of Eq. (\ref{parti}) on $\alpha$ and
$\beta$.


\section{Periodic measurements in the chirality}

In this section we will consider a succession of unitary evolutions
of the QW followed by a process of chirality-and-position
measurement. Let us take the origin as the initial position of the
random walker with the initial condition $|\Psi (0)\rangle $ of Eq.
(\ref{psi0}). The position and chirality of the walker are jointly
measured every $T$ steps, with $T$ sufficiently large in order to
use the asymptotic results for the chirality Eq. (\ref{parti}),
\emph{i.e.} $T\sim2000$. The measurement of any observable of the
system produces the collapse of the wave function into an eigenstate
of the measured operator. Among the several alternatives for
measuring chirality, we choose to measure it in such a way that the
chirality is projected on the ${z}$ direction by the Pauli operator
$\sigma _{z}$, whose eigenstates are the qubit states $|L\rangle
=\binom{1}{0}$ and $|R\rangle =\binom{0}{1}$. The final position can
be any position between $k=-T$ and $k=T$ but of course with
different probabilities. Due to the fact that we are interested in
the time-evolution of the chirality, we can rename the final
position measured as the new position $|0\rangle $, this translation
does not modify the chirality evolution because this QW is
homogeneous in position. Immediately after the measurement has been
performed, the system is in state $|L\rangle |0\rangle $ or
$|R\rangle |0\rangle $ with probability $\Pi _{L}$ or $\Pi _{R}$
respectively.

The system evolves again unitarily during $T$ time steps and then a
second measurement is performed. Let us analyze this second
evolution carefully. Supposing first that the system starts from the
chirality state $|L\rangle$, after $T$ time steps the new GCD is
determined using Eq. (\ref{parti}) with $\alpha =0$, and the value
obtained is $\binom{p}{q}$. If instead we start from the chirality
state $|R\rangle$ once again using Eq. (\ref{parti}) now with
$\alpha =\pi /2$, we obtain a GCD given by $\binom{q}{p}$. In the
general case, we have a probability $\Pi_L$ of starting from the
state $|L\rangle$ and $\Pi_R$ of starting from the state
$|R\rangle$, therefore we must weigh the previous cases by their
probability of occurrence. We obtain in this way the composite
probability expression for the GCD after the second measurement
\begin{equation}
{\left[
\begin{array}{c}
P_{L}(2T) \\
P_{R}(2T)%
\end{array}%
\right] }={\left(
\begin{array}{cc}
p & q \\
q & p%
\end{array}%
\right) }\left[
\begin{array}{c}
P_{L}(T) \\
P_{R}(T)%
\end{array}%
\right] ,  \label{markov2}
\end{equation}%
where
\begin{align}
{p}& =1-\frac{1}{2\sqrt{2}},  \label{pe} \\
{q}& =\frac{1}{2\sqrt{2}},  \label{qu}
\end{align}%
and we define $P_{L}(T)\equiv \Pi _{L}$ and $P_{R}(T)\equiv \Pi
_{R}$.

Eq. (\ref{markov2}) is clearly a master equation and the global
evolution of the system is a Markov process consisting of a
succession of unitary evolutions followed by measurements. Such
processes have the property that for any set of successive times
($T,2T,3T,...,mT$) the conditional probability at $mT$ is uniquely
determined by the value of stochastic variables at $(m-1)T$ and is
not affected by any knowledge of the values at earlier times. In
other words, the future state of the system depends only on the
current state and not on the path of the process. In our case the
measurement of the state of the system is a simple but extreme way
of introducing decoherence that produces a loss of long-range
memory. After performing $m\geq2$ measurements the master equation
becomes
\begin{equation}
{\left[
\begin{array}{c}
P_{L}(mT) \\
P_{R}(mT)%
\end{array}
\right] }={\left(
\begin{array}{cc}
p & q \\
q & p%
\end{array}
\right) }\left[
\begin{array}{c}
P_{L}((m-1)T) \\
P_{R}((m-1)T)%
\end{array}
\right] ,   \label{markov3}
\end{equation}
and the probabilities as functions of the initial GCD evolution are%
\begin{equation}
{\left[
\begin{array}{c}
P_{L}(mT) \\
P_{R}(mT)%
\end{array}
\right] }={\left(
\begin{array}{cc}
p & q \\
q & p%
\end{array}
\right) ^{m-1}}\left[
\begin{array}{c}
\Pi_{L} \\
\Pi_{R}%
\end{array}
\right] ,   \label{markov4}
\end{equation}
where $\Pi_{L}$ and $\Pi_{R}$ are given by Eq. (\ref{parti}). After
some simple algebra, involving the calculation of the eigenvalues of
the previous matrix in Eq. (\ref{markov3}), the general solution of
Eq. (\ref{markov4}) is obtained
\begin{equation}
\left[
\begin{array}{c}
P_{L}(mT) \\
P_{R}(mT)%
\end{array}
\right] =\left(
\begin{array}{cc}
p_{m} & q_{m} \\
q_{m} & p_{m}%
\end{array}
\right) \left[
\begin{array}{c}
\Pi_{L} \\
\Pi_{R}%
\end{array}
\right] ,   \label{solution}
\end{equation}
where
\begin{align}
p_{m} & =\frac{1}{2}\left\{ 1+\left( 2p-1\right) ^{m-1}\right\} ,  \notag \\
q_{m} & =\frac{1}{2}\left\{ 1-\left( 1-2q\right) ^{m-1}\right\}.
\label{coefficients}
\end{align}

\begin{figure}[h]
\begin{center}
\includegraphics[scale=0.4]{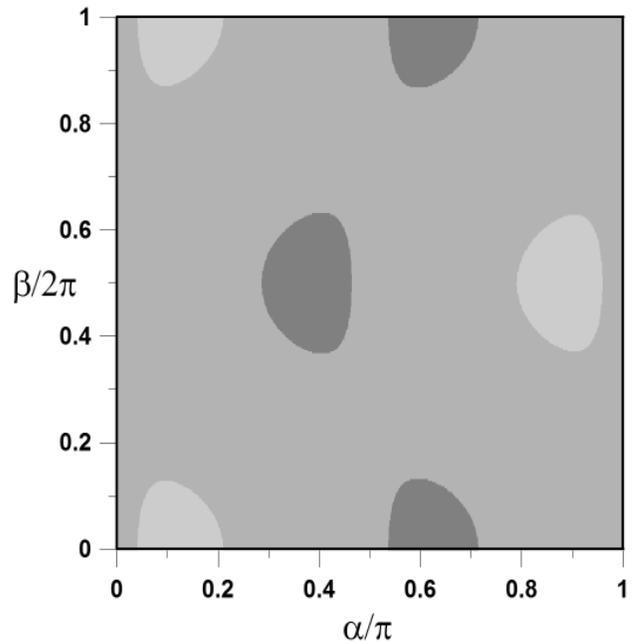}
\end{center}
\caption{The asymptotic probability $P_{R}(2T)$ as a function of
dimensionless angles using Eq. (\protect\ref{solution}) with $m=2$.
Three zones of probability are given with different tonalities of
gray. From the
darker to the lighter zone the intervals of probability are: (1) $\left[0.39,0.45%
\right] $, (2) $\left[0.45,0.52\right]$, (3)
$\left[0.52,0.60\right]$. In section IV this figure also represents
Bob's strategic space for the game with two measurement.} \label{f3}
\end{figure}

In summary the global evolution of the system, throughout a time
interval involving many measurement events, satisfies a master
equation and therefore can be described as a Markovian process. The
system has an unitary evolution only between consecutive
measurements. For a sufficiently large number of measurements, Eq.
(\ref{solution}) implies that both $P_{L}$ and $P_{R}$, at first
sight, tend to $1/2$ independently of the initial conditions. In
order to show this behavior graphically we present Fig. \ref{f3}
where $P_{R}(2T)$ is calculated using the same color code used in
Fig. \ref{f2}. The values of $P_{R}(2T)$ are approximately
restricted to the interval $\left[0.39,0.60 \right]$. This shows
that the interval of probability $\left[0.45,0.52\right]$ has
increased in comparison to the case described by Figs. \ref{f1} and
\ref{f2}. Additionally, we also calculated the probabilities when
three measurements are performed, that is $P_{R}(3T)$ with $m=3$ in
Eq. (\ref{solution}). We verified that $P_{R}(3T)$ is approximately
restricted to the interval $\left[0.45,0.52\right]$, meaning that if
we used once again the color code adopted before, the corresponding
figure would be an uniformly gray square not worthwhile presenting.

\section{Coin-flipping games}
Game theory has been used to explore the nature of quantum
information. Initially quantum games were proposed as a quantum
generalization of their classical counterparts but, due to the
principles of quantum mechanics, new game possibilities have arisen
within this scenario \cite{Meyer,game1Romanelli,zabaleta}. In the
frame of the previous section, let us consider a simple quantum
state flip game played between Alice and Bob. Initially the system
is prepared in the arbitrary state $|\Psi(0)\rangle$ as in Eq.
(\ref{psi0}) where $\alpha$ is chosen by Alice and $\beta$ is chosen
by Bob, or vice versa. Next the QW dynamics evolves with Eq.
(\ref{evolution}) up to the time $t\geq T$. At that time the
chirality of the system is measured. The players agree that the
result of the measurement determines who wins the game. For example,
if the state is $|L\rangle$ Alice wins $\$1$ (Bob loses $\$1$) and
if the state is $|R\rangle$ Bob wins $\$1$ (Alice loses $\$1$). This
is a two-person, zero-sum game; the payoff given to Alice is the
exact opposite of that awarded to Bob. If we suppose rational
players, the election of their strategies is determined by the
election of the initial condition for $|\Psi(0)\rangle$. From Figs.
\ref{f1} and \ref{f2}, which show the dependence of probabilities
$\Pi_{L}$ and $\Pi_{R}$ with $\alpha$ and $\beta$, we can conclude
that the first player to select one of the initial conditions has an
advantage. For example, if Alice chooses first, she needs to choose
$\alpha\in(0.45\pi,0.55\pi)$ (see Fig. \ref{f1}) because in this
way, independently of what Bob chooses, she makes sure that the
probability to obtain $|L\rangle$ is $0.60\leq\Pi_{L}\leq0.65$ in
which case Bob almost certainly loses. On the contrary, if Bob
chooses first (see Fig. \ref{f2}) he needs to choose
$\alpha\in(0,0.1\pi)$ or $\alpha \in(0.9\pi,\pi)$ to obtain a
probability $0.60\leq\Pi_{R}\leq0.71$ independently of $\beta$.
Therefore, the players must make their measurements obeying the
rules of quantum mechanics and the initial conditions of the wave
function determines their strategies. By looking at Figs. \ref{f1}
and \ref{f2}, the players can plan their strategies and quickly
conclude that this is not an equitable game; the first player has
many winning strategies.

A possible variant of the previous game is obtained when a second
measurement at $t\geq2T$ is incorporated. In this new game, after
the first measurement at $t=T$, the system has an unitary evolution
until a second measurement is performed. Once again, Alice wins
$\$1$ and Bob loses $\$1$ if the result of the last measurement is
$|L\rangle$, independently of the result of the previous
measurements. If the last measurement yields $|R\rangle$ Bob wins
$\$1$ and Alice loses $\$1$. Fig.~\ref{f3}, shows the probability
$P_{R}(2T)$ as a function of the initial conditions $\alpha$ and
$\beta$ for the new game. In this figure a large new area with
probability $0.45\leq P_{R}(2T)\leq0.52$ appears. This area shows
that this game is fairer because now the initial
condition $\beta$ can be chosen in order to balance the $P_{L}(2T)$ and $%
P_{R}(2T)$ probabilities. If the number of measurements between unitary
evolutions is increased, we obtain a succession of more equitable games and
in the infinite limit we have a fair game where $P_{L}(\rightarrow%
\infty)=P_{R}(\rightarrow\infty)=1/2$ for all initial conditions.

Taking into account the inequity of the previous games, the players
decide as a strategy to choose $\alpha$ and $\beta$ at random, but
obeying the constraint on $|\Psi(0)\rangle$ given by Eq.
(\ref{psi0}). With the aim to quantify the inequity of these games
we now introduce the mean payoff of the game for each player. In the
first game using Eq. (\ref {parti}) we can reinterpret $\Pi_{L}$ and
$\Pi_{R}$ as the win density for Alice $\sigma_{A}\equiv\Pi_{L}$ and
Bob $\sigma_{B}\equiv\Pi_{R}$ as functions of the parameters
$\alpha$ and $\beta$. Then the probabilities of winning the game are
\begin{align}
\pi_{A} & \equiv\frac{1}{2}\int\limits_{0}^{2\pi}{d\beta}\int\limits_{0}^{%
\pi}\sigma_{A}(\alpha,\beta)\text{ }{d\alpha}=1/2  \label{payoffsilvia} \\
\pi_{B} & \equiv\frac{1}{2}\int\limits_{0}^{2\pi}{d\beta}\int\limits_{0}^{%
\pi}\sigma_{B}(\alpha,\beta)\text{ }{d\alpha=}1-\pi_{A}=1/2,
\label{payoffjuan}
\end{align}
for Alice and Bob respectively. The expected payoff for Alice is $\overset{\_}%
{\pi}_{A}\equiv\$1\pi_{A}-\$1\pi_{B}$ and the expected payoff for Bob is $%
\overset{\_}{\pi}_{B}\equiv\$1\pi_{B}-\$1\pi_{A}$; in this game $\overset{\_}%
{\pi}_{A}=\overset{\_}{\pi}_{B}=\$0$. In the other games, we define
the win densities as in the first game but now using Eq.
(\ref{solution})
\begin{align}
\sigma_{A}& = \left[ 1+\left( 2p-1\right) ^{m-1}\right] \frac{\Pi_{L}}{2}+%
\left[ 1-\left( 1-2q\right) ^{m-1}\right] \frac{\Pi_{R}}{2} ,  \notag \\
\sigma_{B}& = \left[ 1-\left( 1-2q\right) ^{m-1}\right] \frac{\Pi_{L}}{2}+%
\left[ 1+\left( 2p-1\right) ^{m-1}\right] \frac{\Pi_{R}}{2},
\label{alfabeta2}
\end{align}
with $m=2,3,4,...$.

The probabilities of winning the game are calculated using Eq.
(\ref{alfabeta2}) in Eqs.(\ref{payoffsilvia}, \ref{payoffjuan}).
Their values are $\pi_{A}=1/2$, $\pi_{B}=1/2$ again and the expected
payoffs for Alice and Bob are
$\overset{\_}{\pi}_{A}=\overset{\_}{\pi}_{B}=\$0$. Therefore, it is
expected that they will tie on average after many games are played,
provided they choose initial conditions at random.

\section{GCD dynamics with broken links}
\begin{figure}[h]
\begin{center}
\includegraphics[scale=0.3]{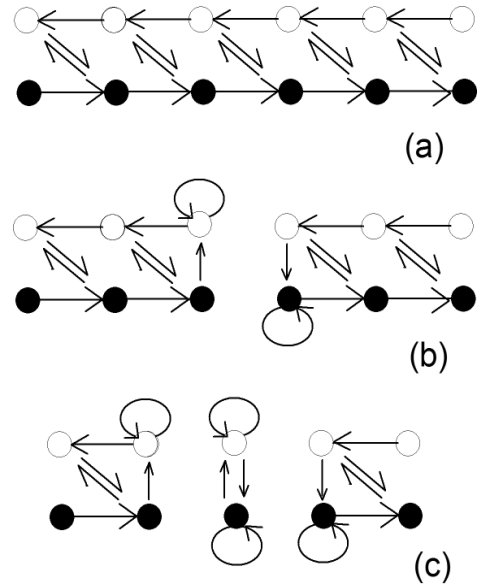}
\end{center}
\caption{ Situations that can arise at a site $k$ of the line when
there are: (a) no broken links (Eq. (\ref{mapa})), (b) one link is
broken (Eqs.(\ref{mapa1},\ref{mapa2})), and (c) both continuous
links are broken (Eq. (\ref{mapa3})). The upper white (lower black)
circles indicate the left (right) chirality amplitude. The arrows
indicate the direction of the probability flux in one time step. At
each circle two arrows point in and two point out in order to
conserve the total probability flux.} \label{f4}
\end{figure}

Let us now consider the breaking of links as a mechanism to
introduce decoherence in the QW \cite{kendon2010}. This model was
presented for the first time in Ref \cite {alejo4}. Suppose that, at
time $t$, a given site $k$ may have one or both of the links
connecting it to its neighboring sites broken. If site $k$ has no
broken links, as shown in Fig. \ref{f4} $a$, then the evolution law
Eq. (\ref{mapa}) is applied. When one or both links at site $k$ are
opened there can be no translation across the broken link and the
evolution must be modified accordingly. If the link to the left of
site $k$ is broken, as shown in Fig. \ref{f4} $b$, the upper
component of the chirality at $k$ receives probability flux from
$k+1$. In order to preserve the flux, the outgoing probability flux
from the upper component at $k$ must be diverted to the lower
component at the same site. The corresponding transformation on the
chirality components is therefore
\begin{align}
a_{k}(t+1) & =a_{k+1}(t)\,\cos\theta\,+b_{k+1}(t)\,\sin\theta,\,  \notag \\
b_{k}(t+1) & =a_{k}(t)\,\sin\theta\,+b_{k}(t)\,\cos\theta.  \label{mapa1}
\end{align}
If the broken link is to the right of $k$, the situation is similar
and the transformation is
\begin{align}
a_{k}(t+1) & =a_{k}(t)\,\cos\theta\,-b_{k}(t)\,\sin\theta,\,  \notag \\
b_{k}(t+1) & =a_{k-1}(t)\,\sin\theta\,-b_{k-1}(t)\,\cos\theta.  \label{mapa2}
\end{align}
Finally, if site $k$ is isolated, as in Fig. \ref{f4} $c$, the
unitary operation is followed by a chirality exchange,
\begin{align}
a_{k}(t+1) & =a_{k}(t)\,\cos\theta\,-b_{k}(t)\,\sin\theta,\,  \notag \\
b_{k}(t+1) & =a_{k}(t)\,\sin\theta\,+b_{k}(t)\,\cos\theta.  \label{mapa3}
\end{align}
Hence, there are four situations that can arise at a site $k$: (a)
no link is broken, (b) the link on the left side is broken, (c) the
link on the right side is broken, and (d) both links are broken. The
global evolution of the system depends on the application of one of
the four maps in each spatial position, where each map has
associated its correlative operator. Thus the dynamical evolution of
the GCD can be characterized by four unitary operators which will be
called in generic terms $U_n$ with $n=0,1,2,3$ where $U_0$ will be
related to the operator $U$ of Eq. (\ref{evolution}). A statistical
description can be obtained combining the operators into a single
evolution equation with the appropriate statistical weights.

To implement the algorithm of the quantum walk with broken links we
proceed as follows. At each time step $t$, the state of the links in
the line is defined. Each link has a probability $r $ of breaking in
a given time step, being $r$ the only parameter in the model.
Therefore the probability that a given site has no adjacent broken
link is
\begin{align}
r_0=(1-r)^2,  \label{r0}
\end{align}
that it has a left or right broken link is
\begin{align}
r_1=r_2=r(1-r),  \label{r1}
\end{align}
and that it is isolated is
\begin{align}
r_3=r^2.  \label{r3}
\end{align}
These probabilities are appropriate statistical weights since they satisfy
\begin{align}
r_0+r_1+r_2+r_3=1.  \label{norm}
\end{align}

In order to study the chirality dynamics of the QW we introduce the
reduced density operator
\begin{align}
\rho_{c}=\mathrm{tr}\rho ,  \label{rhoc}
\end{align}
where the operator `$\mathrm{tr}$' is the partial trace taken over the
positions and $\rho$ is density matrix of the quantum system
\begin{align}
\rho=|\Psi (t)\rangle \langle \Psi (t)|.  \label{rho}
\end{align}
Using the wave function given by Eq. (\ref{spinor}) and its
normalization properties, the reduced density matrix obtained in
Ref. \cite{Carneiro} is
\begin{equation}
\rho_{c} =\left(
\begin{array}{cc}
P_{L}(t) & Q(t) \\
Q(t)^{\ast } & P_{R}(t)%
\end{array}%
\right) .  \label{rho}
\end{equation}
This matrix depends explicitly on the GCD and its time evolution is
given by the time evolution of the GCD. The dynamical equation for
$\rho_{c}(t)$ is obtained using the density matrix formalism
\cite{Kraus,Brun,Romanelli09k}
\begin{equation}
\rho_{c}(t+1)=\sum_{n=0}^{3}A_{n}\rho_{c}(t)A_{n}^{\dag },  \label{map}
\end{equation}
where a set of Kraus operators ($\{{A_{n}}\}$ with $n=0,1,2,3$) has been
introduced in order to simulate the decoherence produced by the broken-links
process. The Kraus operators are defined by
\begin{equation}
{A_{n}\equiv\sqrt{r_n} U_n},  \label{kraus}
\end{equation}
and, to preserve the trace of the density matrix, they satisfy
\begin{equation}
\sum_{n=1}^{N}A_{n}A_{n}^{\dag }=I,  \label{kraus}
\end{equation}
where $I$ is the identity matrix.

We want to study the quantum evolution of Eq. (\ref{map}), but this
task is not trivial since we do not know the set of operators
$\{{A_{n}}\}$ explicitly. However, we can infer some knowledge about
the quantum evolution of the GCD  if we follow the procedure
developed in section $II$ to obtain Eqs. (\ref{master})
from Eq. (\ref{mapa}). Starting from the maps Eqs.(\ref{mapa},\ref{mapa1},%
\ref{mapa2},\ref{mapa3}) the corresponding dynamical equations for the GCD
are built. Next these equations are weighted with the probabilities $r_n$
to finally obtain the master equation for the averaged GCD
\begin{align}
{\left[
\begin{array}{c}
P_{L}(t+1) \\
P_{R}(t+1)%
\end{array}
\right] } & ={\left(
\begin{array}{cc}
\cos^{2}\theta & \sin^{2}\theta \\
\sin^{2}\theta & \cos^{2}\theta%
\end{array}
\right) }\left[
\begin{array}{c}
P_{L}(t) \\
P_{R}(t)%
\end{array}
\right].   \label{master1}
\end{align}
This equation has no dependence with the parameter $r$ and, in
opposition to Eq. (\ref{master}), has a negligible interference term
$Q(t)$ due to the random phases introduced by the lack of coherence
between the terms of the sum that define the interference in Eq.
(\ref{inter}). Therefore, if $Q(t)$ is negligible it is clear that
Eq. ( \ref{map}) is equivalent to Eq. (\ref{master1}), and we could
use this equivalence to obtain information about the operators
$\{{A_{n}}\}$.
\begin{figure}[h]
\begin{center}
\includegraphics[scale=0.4]{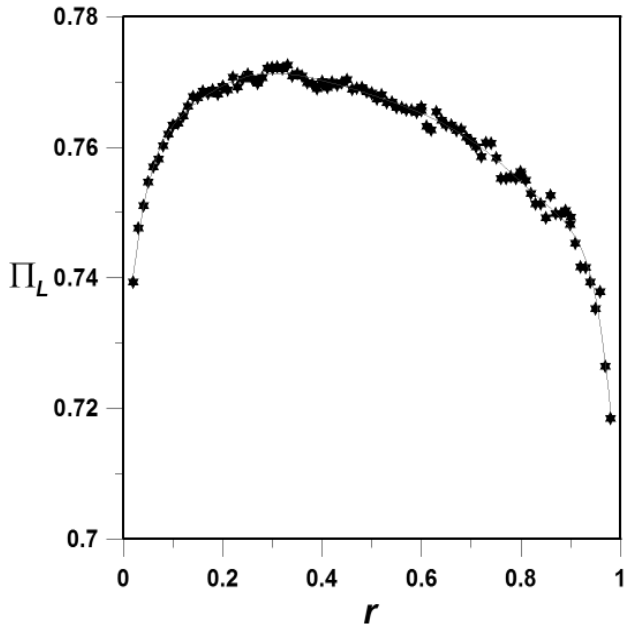}
\end{center}
\caption{The asymptotic probability $\Pi_{L}$ as a function of the
probability $r$ with $\protect\theta=\protect\pi/4$ and the initial
condition $\frac{1}{\sqrt{2}}\binom{1}{1}$. The stars are the values
obtained from the numerical calculation and the thin line is a
polynomial adjustment. The calculation has been made using the map Eq. (%
\protect\ref{mapa}) for the left side of the evolution and the maps Eqs. (%
\protect\ref{mapa},\protect\ref{mapa1},\protect\ref{mapa2},\protect\ref%
{mapa3}) weighted with the values given by Eqs.
(\protect\ref{r0},\protect \ref{r1},\protect\ref{r3}) for the right
side of the evolution.} \label{f5}
\end{figure}

\begin{figure}[h]
\begin{center}
\includegraphics[scale=0.4]{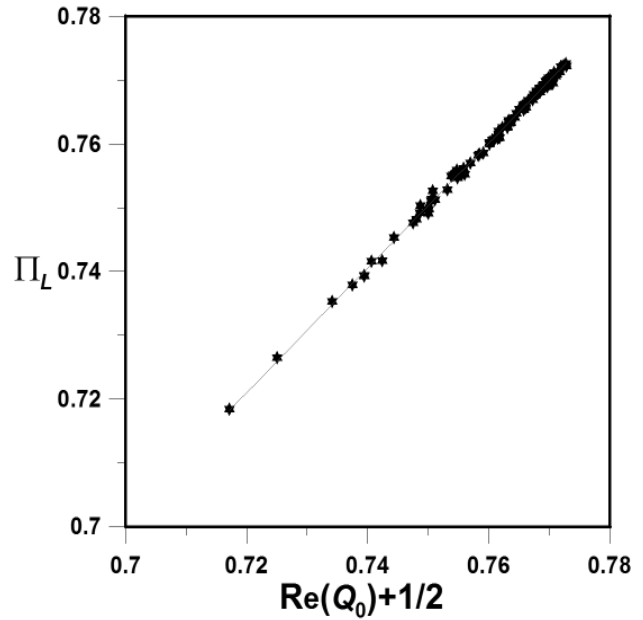}
\end{center}
\caption{The asymptotic probability $\Pi_{L}$ as a function of the
dimensionless term Re$(Q_0)+1/2$, with
$\protect\theta=\protect\pi/4$ (see Eq. (\protect\ref{estacio})) and
the initial condition $\frac{1}{\sqrt{2}}\binom{1}{1}$. The stars
are the values obtained from the numerical calculation and the thin
line is a linear adjustment with
slope $1$. The calculation has been made using the map Eq. (\protect\ref%
{mapa}) for the left side of the evolution and the maps Eqs. (\protect\ref%
{mapa},\protect\ref{mapa1},\protect\ref{mapa2},\protect\ref{mapa3}) weighted
with the values given by Eqs. (\protect\ref{r0},\protect\ref{r1},\protect\ref%
{r3}) for the right side of the evolution.}
\label{f6}
\end{figure}
The stationary solution of Eq. (\ref{master1}) is $\Pi_{L}=\Pi_{R}=1/2$
independently of the initial conditions and of $\theta$. We implemented a
numerical code with the four maps Eqs. (\ref{mapa},\ref{mapa1},\ref{mapa2},%
\ref{mapa3}), weighted with the values $r_0$, $r_1$, $r_2$, $r_3$
respectively, and have verified the asymptotic value $\Pi_{L}=1/2$ for several
initial conditions and several values of $\theta$ and $r$. Thus, the
theoretical result given by Eq. (\ref{master1}) has been numerically confirmed.
\begin{figure}[h]
\begin{center}
\includegraphics[scale=0.4]{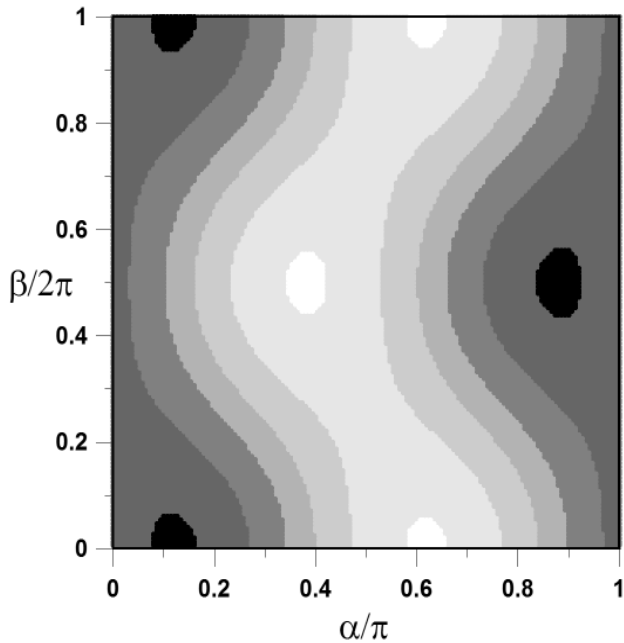}
\end{center}
\caption{ The asymptotic probability $\Pi_{L}$ as a function of
dimensionless angles. The calculation has been made using the map Eq. (%
\protect\ref{mapa}) for the left side of the evolution and the maps Eqs. (%
\protect\ref{mapa},\protect\ref{mapa1},\protect\ref{mapa2},\protect\ref%
{mapa3}) weighted with the values given by Eqs.
(\protect\ref{r0},\protect \ref{r1},\protect\ref{r3}) for the right
side of the evolution, with $r=0.3$. Seven zones of probability are
given with different tonalities of gray, from the black to the white
zone the intervals of probability are: (1)
$\left[0.738,0.746\right]$, (2) $\left[0.746,0.755\right]$, (3)
$\left[0.755,0.763\right]$, (4) $\left[0.763,0.771\right]$, (5)
$\left[0.771,0.780\right]$, (6) $\left[0.780,0.789\right]$, (7)
$\left[0.789,0.797\right]$.} \label{f7}
\end{figure}

The usual QW dynamics, Eq. (\ref{mapa}), has the property of spreading
over the line quadratically faster than its classical analog and
after a long time its wave-function is very dispersed on the line.
This situation makes it interesting to study the decoherence when
only some segment on the line is affected by the breakage of links. Such decoherence mechanisms may be
relevant as a model of experimental realizations in quantum computation \cite%
{Exp,ProExp,Berman} as well as in any problem related to the
transfer of quantum information through space using QW
\cite{childs,Lovett}. In order to investigate this type of
decoherence, we propose a composed non-homogeneous QW with two
different dynamics starting
form the origin: for the left half-line the evolution is determined by Eq. (\ref%
{mapa}) and for the right half-line it is determined by the set of
equations Eqs. (\ref{mapa},\ref{mapa1},\ref{mapa2},\ref{mapa3}),
weighted with the values $r_0$, $r_1$, $r_2$, $r_3$ respectively.
The stochastic behavior of the GCD , for an ensemble of these QWs,
will depend, for the left side, on the usual QW and, for the right
side, on the QW with broken links. Therefore the right half-line
chirality distribution satisfies Eq. (\ref{master1}) and the left
half-line chirality distribution satisfies Eq. (\ref{master}). From
this reasoning, it is clear that the GCD is the sum of the above
right and left partial distributions and it satisfies Eq.
(\ref{master}), where the interference term emerges from the
amplitudes of the wave function belonging to the left side of the
evolution. However, it should be noted that the incorporation of the
interference term of the right side of the evolution does not modify
this equation because it remains negligible.

The evolution of the system ceases to be unitary owing to
decoherence, however in spite of the stochastic nature of the system
Re$(Q_0)\neq0$ and Eq. (\ref{estacio}) maintains the dependence
between the asymptotic values $\Pi_L$ and Re$(Q_0)$. The system
loses its integrability and the explicit value of $\Pi_L$ is to be
calculated numerically. We have performed this calculation, as a
function of the probability $r$ with $\theta= \pi/4$ using the
original maps Eqs. (\ref{mapa},\ref{mapa1},\ref{mapa2},\ref{mapa3})
with an ensemble of $100$ dynamical trajectories and $2000$ time
steps. Figure \ref{f5} shows that $\Pi_L$ has a parameter dependence
on $r$. Figure \ref{f6} shows the linear dependence between $\Pi_L$
and Re$(Q_0)+1/2$, with slope $1$ for $\theta=\pi/4$, verifying the
stationary solution given by Eq. (\ref{estacio}). Therefore, we
conclude that there is a perfect agreement between the theoretical
stochastic approach presented in this section and the numerical
calculation using the original maps.
%aca me saque las comas por sugerencia de Victor.
We also found from the numerical calculation that the asymptotic
distribution $\Pi_L$ depends on the initial conditions. Figure
\ref{f7} presents the numerical calculation of the distribution
$\Pi_{L}$, with $3000$ time steps and the initial condition given by
Eq. (\ref{psi0}).

Comparing Fig. \ref{f7} with Fig. \ref{f1}, it is possible to
distinguish several similarities, such as the number of light and
dark spots in both figures. These similarities, show that the
asymptotic behavior depends on the initial condition. In other
words, this type of decoherence erases the unitary behavior of the
system but does not completely erase its dependence on the initial
condition. However, the range of values of the probability is now
restricted to $\sim\left[0.7,0.8\right]$ while in the Fig. \ref{f1}
this range was $\sim\left[0.3,0.7\right]$. Such a decrease in the
range shows that the trend of the system is to go to a single
asymptotic distribution that will only depend on the parameter $r$,
as seen in Fig. \ref{f5}.

\section{Conclusions}

\label{sec:conclusion}

Decoherence in quantum systems like the QW has been extensively
studied. Analytical and numerical results on the effect of different
kinds of noise have shown that quantum properties are highly
sensitive to random events. In this paper, we studied the asymptotic
behavior of the QW on the line when it is subjected to different
sources of decoherence. We focused on the dynamics of the GCD when
decoherence is introduced in two different ways. In the first case
decoherence is introduced through periodic measurements of position
and chirality. It has been shown that the evolution of the GCD, in a
time scale involving many measurements, can be described as a
Markovian process and its master equation has been obtained
analytically. This  master equation allows us to build new quantum
games, similar to the usual spin-flip game, where as a novelty the
players perform measurements on the QW system. These games are
characterized by a continuous space of strategies, and the selection
of an initial condition determines the particular strategy chosen by
the player. These games may be used as a tool to study quantum
algorithms subjected to external decoherence, as in the extreme case
of measurement.

In the second case, the decoherence is introduced through the
influence of randomly broken links affecting the QW dynamics. Two
different variants of this second case have been investigated. These
are: (a) All the QW links have a certain probability of being
broken, the GCD is described by a Markovian process.  Its master
equation is obtained analytically and it is shown to be equivalent
to the dynamical equation of the density matrix. It was shown that
the dynamics is independent both of the breakage probability and of
the coin bias of the QW. (b) Only the links on one of the half-lines
(say the right one) are affected by the possibility of breakage. In
this case the dynamical evolution of the GCD is not described by a
Markovian process but it has an asymptotic value that depends on the
initial conditions and the breakage probability. Here the
interference term $Q_0$ is not negligible but an asymptotic
stationary relation between $Q_0$ and the GCD has been analytically
found. The behavior of this composite QW is at first sight
unexpected since usually decoherence destroys the unitary
correlation, providing a route towards a classical-like behavior
described by a Markov process. To understand this behavior suppose a
localized initial condition far to the left of the origin, situated
thus in a zone where the evolution is determined by the usual QW
map. As time passes, the wave function spreads in both directions
and, due to the dynamical evolution of the map, the extreme left
branch of the wave function will never be influenced by the
existence of decoherence introduced into the right branch of the
wave function. In this way, the wave function keeps some information
about the initial condition. We conclude that the effect of
decoherence of the second kind studied in this work does not
necessarily transform the quantum system into a dissipative one such
as a Markov process. In more general terms, the mere presence of
noise does not assure the transition of a quantum system to
classicality.

We acknowledge the comments made by V. Micenmacher and the support
from PEDECIBA and ANII.

\end{document}